\documentclass[titlepage,prd,byrevtex,amsmath,
preprint,
preprintnumbers,
tightenlines,
]{revtex4-1}

\usepackage{graphicx}
\usepackage{bm}
\usepackage{hyperref}
\newcommand{\beq}{\begin{equation}}
\newcommand{\eeq}{\end{equation}}
\newcommand{\eq}[1]{Eq.~(\ref{#1})}

\begin{document}

\title{Universal Binding and Recoil Corrections to Bound State $g$-Factors in Hydrogenlike Ions}
\author {Michael I. Eides}
\altaffiliation[Also at ]{Petersburg Nuclear Physics Institute,
Gatchina, St.Petersburg 188300, Russia}
\email{eides@pa.uky.edu, eides@thd.pnpi.spb.ru}
\author{Timothy J. S. Martin}
\email{tjmart1@uky.edu}
\affiliation{Department of Physics and Astronomy,
University of Kentucky, Lexington, KY 40506, USA}
%\date{\today}

\begin{abstract}

The leading relativistic and recoil corrections to bound state $g$-factors of particles with arbitrary spin are calculated. It is shown that these corrections are universal for any spin and depend only on the free particle gyromagnetic ratios. To prove this universality we develop nonrelativistic quantum electrodynamics (NRQED) for charged particles with an arbitrary spin. The coefficients in the NRQED Hamiltonian for higher spin particles are determined only by the requirements of Lorentz invariance and local charge conservation in the respective relativistic theory. For spin one charged particles the NRQED Hamiltonian follows from the renormalizable QED of the charged vector bosons. We show that universality of the leading relativistic and recoil corrections can be explained with the help of the Bargmann-Michael-Telegdi equation.

\end{abstract}

%\pacs{12.20.-m,12.20.Ds,13.40.Em,31.15.aj,31.30.J-,31.30.js}
%\keywords{gyromagnetic ratio}

\preprint{UK-10-03}

\maketitle

Gyromagnetic ratios of particles in hydrogenlike bound states have become in the last ten-fifteen years an active field of experimental and theoretical research. The gyromagnetic ratio of a bound electron is proportional to the ratio of the spin flip and cyclotron frequencies of a hydrogenlike ion and to the electron-ion mass ratio. The experimental uncertainties of the ratio of the spin flip and cyclotron frequencies of the hydrogenlike carbon $^{12}C^{5+}$ and oxygen $^{16}O^{7+}$ were reduced to $5-7$ parts in $10^{10}$, see \cite{haffner2003,verdu2004} and review in \cite{mtn2008}. The theoretical expression for the bound state $g$-factor was also greatly improved recently (see, e.g. \cite{pcjy,jent2009} and references in \cite{mtn2008}), and the theoretical uncertainty was reduced to $1.5-5.5$ parts in $10^{11}$. As a result measurements of the bound electron $g$-factor became the best source for precise values of the electron mass in atomic units \cite{mtn2008}. This bright picture is marred by the discrepancy on the magnitude of the leading relativistic and recoil corrections to bound state $g$-factors existing in the literature \cite{eg,mfjetp,fmyaf}. This discrepancy shifts the theoretical value of the bound state $g$-factors of the hydrogenlike carbon $^{12}C^{5+}$ and oxygen $^{16}O^{7+}$ by about $2-3$ parts in $10^{11}$. It will become even more phenomenologically relevant if proposed improvement \cite{quint2008} of the experimental accuracy by two orders of magnitude is achieved. Theoretically, discrepancy between different results for the leading relativistic and recoil corrections to bound state $g$-factors is connected with different treatments of the spin dependence of these corrections. Below we derive an effective NRQED Hamiltonian for charged particles with arbitrary spins and calculate the leading relativistic and recoil corrections to the bound state $g$-factors in loosely bound two-particle systems. We show that these corrections are universal for all spins; they do not depend on the magnitude of spin.

A loosely bound two-particle system is effectively nonrelativistic, with characteristic velocities of constituents of order $Z\alpha$. We are looking for the leading binding and recoil corrections of order $(Z\alpha)^2$. NRQED is a natural framework for calculation of these corrections. We first consider leading nonrecoil corrections to the gyromagnetic ratio of order $(Z\alpha)^2$. To calculate all such corrections we need the NRQED Lagrangian that includes all terms in nonrelativistic expansion up to and including $v^2$. We should also include in the effective Lagrangian terms with the external Coulomb field $A_0$, since for such field $\langle eA_0\rangle\sim(Z\alpha)^2$. The NRQED Lagrangian for spin one half case is well known (see, e.g., \cite{kn1}). The coefficients in the NRQED Lagrangian for charged particles with an arbitrary spin should be determined from comparison with the results of the respective relativistic theory. The problem is that renormalizable QED for charged particles with high spin does not exist. The rules for calculation of all one-photon interactions of charged particles with arbitrary spin were constructed some time ago in \cite{kms_pl,kms_jetp}. This construction uses only Lorentz invariance and local current conservation, and it should be valid for charged particles of arbitrary spin. The interaction vertex in the approach of \cite{kms_pl,kms_jetp} is a direct generalization of the ordinary spin one half vertex

\beq \label{vertex}
\Gamma_\mu=e\frac{(p_1+p_2)_\mu}{2m}F_e(q^2,\tau)
-F_m(q^2,\tau)\frac{e\Sigma_{\mu\nu}q^\nu}{2m},
\eeq

\noindent
where $q=p_2-p_1$, $\Sigma_{\mu\nu}$ is the generalization of ordinary spin one half $\sigma_{u\nu}$,  $S_\mu$ is a covariant spin four-vector, $\tau=(q\cdot S)^2$, and $F_e(0,0)=1$, $F_m(0,0)=g/2$. The wave functions are spinors with dotted and undotted indices that are symmetrized among themselves (for more details see \cite{kms_pl,kms_jetp,blp}). The form of the vertex in \eq{vertex} is uniquely fixed by the requirements of Lorentz invariance, $C$, $P$ and charge conservation. Charged particles with higher spins automatically carry higher multipole moments that arise as coefficients in expansion of the form factors $F_e$ and $F_m$ over $\tau$. These intrinsic electric and magnetic multipole moments are treated phenomenologically, and we do not try to calculate them. The phenomenological approach to multipole moments is an advantage for our purposes because we would like to describe how $g$-factors of not necessarily electromagnetic origin (for example the $g$-factor of a spin one deuteron) change in a loosely bound electrodynamic system.

In the spin one half case the NRQED Lagrangian is constructed from the gauge invariant operators $\bm D=\bm\nabla-ie\bm A=i(\bm p-e\bm A)$, $\bm E$, $\bm B$, and $\bm S$. For higher spin particles, besides the spin operator, we should also include higher irreducible intrinsic multipole moments as the building blocks of the NRQED Lagrangian. Technically these  multipole moments are polynomials in the components of the spin operator that for higher spins do not reduce to numerical tensors and operators linear in spin. The most general NRQED Lagrangian has the form (compare with \cite{kn1})

%\begin{widetext}
\begin{eqnarray} \label{lagr}
{\cal L}=\phi^+\Biggl\{i(\partial_0+ieA_0)+\frac{\bm D^2}{2m}+\frac{\bm D^4}{8m^3}+c_F\frac{e{\bm S}\cdot{\bm B}}{2m}
+c_D\frac{e({\bm D}\cdot{\bm E}-{\bm E}\cdot{\bm D})}{8m^2}
\nonumber
\\
+c_Q\frac{eQ_{ij} (D_iE_j-E_iD_j)}{8m^2}
+c_S\frac{ie{\bm S}\cdot({\bm D}\times{\bm E}-{\bm E}\times{\bm D})}{8m^2}
+c_{W1}\frac{e[{\bm D}^2({\bm S}\cdot{\bm B})+({\bm S}\cdot{\bm B}){\bm D}^2]}{8m^3}
\nonumber
\\
+c_{W2}\frac{-e D^i({\bm S}\cdot{\bm B})D^i}{4m^3}
+c_{p'p}\frac{e[({\bm S}\cdot{\bm D})({\bm B}\cdot{\bm D})
+({\bm D}\cdot{\bm B})({\bm S}\cdot{\bm D})]}{8m^3}+\dots\Biggr\}\phi,
\end{eqnarray}
%\end{widetext}

\noindent
where $Q_{ij}=S_iS_j+S_jS_i-(2/3)\bm S^2\delta_{ij}$ is proportional to the electric quadrupole moment operator ($Q_{ij}\equiv0$ for spin one half), and  $\phi$ is a $2S+1$-component spinor field for a particle with spin $s$. We included in the Lagrangian in \eq{lagr} operators of dimensions not higher than four, except those (like the terms with derivatives of magnetic field) that are irrelevant for calculation of the leading recoil corrections. Let us mention that gauge invariant bilinears in $\bm E$ and $\bm B$ are of too high order to generate leading relativistic contributions of order $(Z\alpha)^2$ to bound state $g$-factors.

The coefficients in \eq{lagr} are usually determined from comparison of the one- and two-photon scattering amplitudes in NRQED and relativistic QED. Although some terms in \eq{lagr} are bilinear in $\bm A$ and $\bm E$ all such terms can be restored from one-photon terms due to gauge invariance. Then the one-photon relativistic vertex in \eq{vertex} is sufficient for calculation of all the coefficients in \eq{lagr}. We calculated scattering amplitudes off an external electromagnetic field using the nonrelativistic Lagrangian in \eq{lagr} and using the relativistic one-photon vertex in \eq{vertex} at $\tau=0$. In the relativistic calculation we used noncovariantly normalized particle spinors in the generalized standard representation, which is necessary for consistency with the respective nonrelativistic results. Diagrammatically this choice of spinors and representation corresponds to the Foldy-Wouthuysen transformation (for more details, see, e.g., \cite{blp}). After nonrelativistic expansion we compared results of the relativistic calculation with the nonrelativistic ones and obtained values of all constants in the Lagrangian in \eq{lagr}

\begin{eqnarray}
\label{coeff}
c_F=\frac{g}{2}, c_D=(g-1)\frac{\bm\Sigma^2}{3}, c_S=g-1,  c_Q=-2\lambda(g-1),
\nonumber\\
c_{W1}=\frac{g+2}{4},\quad c_{W2}=\frac{g-2}{4},\quad c_{p'p}=\frac{g-2}{2},\qquad
\end{eqnarray}

\noindent
where $\bm\Sigma^2=4S$, $\lambda=1/(2S-1)$ for integer spin and $\bm\Sigma^2=4S+1$, $\lambda=1/(2S)$ for half integer spin. Dependence on the magnitude of charged particle spin arose in the coefficients before the Darwin term and the induced electric quadrupole interaction. The $g$-factor in \eq{vertex}, \eq{coeff} is the total gyromagnetic ratio of a free nonrelativistic particle defined through the effective interaction Hamiltonian $H_{int}=-ge/(2m)\bm B\cdot\bm S$. If the charged particle is subject only to electromagnetic interaction then $g$ reduces to a sum of the QED perturbation series. For spin one half the coefficients in \eq{coeff}  coincide with the respective coefficients in \cite{kn1}, if the phenomenological $g$-factor is substituted in the expressions in \cite{kn1} instead of the perturbative $g=2(1+\alpha/2\pi)$. As an independent test of the effective Lagrangian in \eq{lagr} we considered the charged $W^\pm$-boson sector of the Glashow-Weinberg-Salam Electroweak Theory amended by the anomalous magnetic moment term. We derived the effective NRQED Lagrangian for the $W^\pm$ bosons. This Lagrangian coincides with the Lagrangian in \eq{lagr} for spin one charged particles.

The NRQED coefficients in \eq{coeff} are calculated ignoring all relativistic loop diagrams and $q^2$ and $\tau$ dependence of the form factors in \eq{vertex}. Both the loop diagrams in relativistic QED and multipole expansion of the form factors would generate further corrections to the coefficients in \eq{coeff}. However, we are interested only in corrections to bound state $g$-factors of order $(Z\alpha)^2\sim v^2$. The coefficients in the effective Lagrangian are calculated comparing  scattering amplitudes in relativistic and nonrelativistic theories. Counting of powers of the coupling constants in the case of scattering amplitudes calculated at a generic kinematical point is trivial. In ordinary renormalizable spin one half QED (as well as in the renormalizable QED of spin one $W^\pm$ vector bosons) all diagrams, besides those that give contributions only to the free particle $g$-factors, generate corrections to the coefficients that are additionally suppressed by powers of $Z\alpha$. We expect the same effect in any reasonable theory for higher spin particles. It is also obvious that accounting for $q^2$ and $\tau$ dependent terms in the form factors in \eq{vertex} generates terms suppressed by additional powers of $Z\alpha$. We do not need to consider two-photon Compton effect diagrams, since all terms in \eq{lagr} bilinear in fields can be restored from one-photon diagrams with the help of gauge invariance. Any gauge invariant terms connected with the two-photon diagrams are of too high order in $Z\alpha$ to contribute to the leading relativistic corrections of order $(Z\alpha)^2$. Hence, the Lagrangian in \eq{lagr} with the coefficients from \eq{coeff} is sufficient for calculation of the leading relativistic corrections to the bound $g$-factor in the nonrecoil case.

Our goal is also to calculate recoil corrections of order $(Z\alpha)^2$ that are linear and quadratic in the mass ratio. To this end we need to construct an effective two-particle NRQED Hamiltonian for a loosely bound electrodynamic system of two particles.  The interaction between two charged particles with accuracy up to $(Z\alpha)^2$ is described by the one photon exchange which generates Coulomb and Breit interactions. We calculated the one-photon potential for two particles with arbitrary spins and magnetic moments and obtained

%\begin{widetext}
\begin{eqnarray} \label{breit}
V_{int}({\bm p}_1,{\bm p}_2,{\bm r})
=e_1e_2
\biggl[\frac{1}{4\pi r}-(g_1-1) \frac{1}{8m_1^2}\frac{\bm\Sigma_1^2}{3}\delta({\bm r})
-(g_1-1)\frac{3\lambda_1}{\pi}\frac{r^ir^jQ^{(1)}_{ij}}{16m_1^2 r^5}
\nonumber\\
-(g_2-1)\frac{1}{8m_2^2}\frac{\bm\Sigma_2^2}{3}\delta({\bm r})
-(g_2-1)\frac{3\lambda_2}{\pi}\frac{r^ir^jQ^{(2)}_{ij}}{16m_2^2r^5}
-\frac{{\bm r}({\bm r}\cdot{\bm p}_1)\cdot{\bm p}_2}{8\pi m_1m_2r^3}-\frac{{\bm p}_1\cdot{\bm p}_2}{8\pi m_1m_2r}
\nonumber\\
-(g_1-1)\frac{2{\bm S}_1\cdot({\bm r} \times{\bm p}_1)}{16 \pi m_1^2r^3}
+g_1\frac{2{\bm S}_1\cdot({\bm r}\times{\bm p}_2)}{16\pi m_1m_2r^3}
+(g_2-1)\frac{2{\bm S}_2\cdot({\bm r}\times{\bm p}_2)}{16 \pi m_2^2r^3}
\nonumber\\
-g_2\frac{2{\bm S}_2\cdot({\bm r}\times{\bm p}_1)}{16\pi m_1m_2r^3}
+\frac{g_1g_2}{16\pi m_1m_2}\left(\frac{{\bm S}_1\cdot{\bm S}_2}{r^3}
-\frac{3({\bm S}_1\cdot{\bm r})({\bm S}_2\cdot{\bm r})}{r^5}-\frac{8\pi}{3}{\bm S}_1\cdot{\bm S}_2\delta({\bm r})\right)\biggr],
\end{eqnarray}
%\end{widetext}

\noindent
where $\bm r_{1(2)}$, $\bm p_{1(2)}$, $\bm S_{1(2)}$, $m_{1(2)}$, $g_{1(2)}$, and $Q^{(1(2))}_{ij}$ are the coordinate, momentum, spin, mass, gyromagnetic ratio, and induced quadrupole moment of the first (second) particle, and  $\bm r=\bm r_1-\bm r_2$ is the relative coordinate.

This interaction is a natural generalization of the spin one half one-photon potential (see, e.g., \cite{blp}). The only difference is that like in the Lagrangian in \eq{lagr} the coefficients in \eq{breit} before the Darwin terms depend on the magnitude of particles' spins, and new terms with electric quadrupole moments arise. The interaction in \eq{breit} is calculated in the absence of the external magnetic field that is present in the $g$-factor problem. This drawback is easily repaired by the minimal substitution $\bm p_i\to\bm p_i-e_i\bm A_i$, $\bm A_i=\bm B\times \bm r_i/2$.

Combining the nonrecoil Lagrangian in \eq{lagr} and the one-photon potential (after minimal substitution) in \eq{breit} we obtain a total effective two-particle NRQED Hamiltonian for electromagnetically interacting particles with arbitrary spins (we preserve only the terms relevant for calculation of the $g$-factor contributions)

\beq \label{totham}
H=H_1+H_2+H_{int},
\eeq

\noindent
where

%\begin{widetext}
%\beq
\begin{eqnarray}
H_1=
\frac{({\bm p}_1-e_1{\bm A}_1)^2}{2m_1}
-g_1\frac{e_1}{2m_1}({\bm S}_1\cdot{\bm B})(1-\frac{\bm p_1^2}{2m_1^2})
-(g_1-2)\frac{e_1}{2m_1}({\bm S}_1\cdot{\bm B})\frac{{\bm p}_1^2}{2m_1^2}
\nonumber\\
+(g_1-2)\frac{e_1}{2m_1}\frac{({\bm p}_1\cdot{\bm B})({\bm S}_1\cdot{\bm
p}_1)}{2m_1^2},
%\eeq
\end{eqnarray}
\begin{eqnarray} \label{trexch}
H_{int}=
\frac{e_1e_2}{4\pi r}
+e_1e_2\biggl[-(g_1-1)\frac{2{\bm S}_1\cdot({\bm r} \times({\bm p}_1-e_1\bm
A_1))}{16 \pi m_1^2r^3}
+g_1\frac{2{\bm S}_1\cdot({\bm r}\times({\bm p}_2-e_2\bm A_2))}{16\pi
m_1m_2r^3}
\nonumber\\
+(g_2-1)\frac{2{\bm S}_2\cdot({\bm r}\times({\bm p}_2-e_2\bm A_2))}{16 \pi
m_2^2r^3}
-g_2\frac{2{\bm S}_2\cdot({\bm r}\times({\bm p}_1-e_1\bm A_1))}{16\pi
m_1m_2r^3}\biggr],
\end{eqnarray}
%\end{widetext}

\noindent
and $H_2$ is obtained from $H_1$ by the substitution $1\to2$.

The nonrelativistic effective two-particle Hamiltonian describes all (nonrecoil and recoil) leading relativistic corrections to bound state $g$-factors of each of the constituents. To calculate these corrections we need to separate effects of the bound system motion as a whole from the internal effects. This task is not quite trivial because the center of mass variables do not separate in the presence of external field. For the current case of a small magnetic field a solution was suggested in \cite{gh,eg}. The main idea is to insist that the center mass of a loosely bound system moves in an external field exactly in the same way as a respective elementary particle with the same mass and charge. To satisfy this transparent physical requirement transition to the standard center of mass coordinates $\bm r=\bm r_1-\bm r_2$, $\bm R=\mu_1\bm r_1+\mu_2\bm r_2$, $\mu_i=m_i/(m_1+m_2)$  should be accompanied by the unitary transformation of the Hamiltonian  $H\to U^{-1}HU$, where $U=e^{i(e_1\mu_2-e_2\mu_1)A(\bm R)\cdot\bm r}$. After this transformation we extract from the transformed Hamiltonian terms that describe the spin interaction with an external field. The final Hamiltonian for the first particle is

\begin{widetext}
\begin{eqnarray}
H^{(1)}_{spin}
=-\frac{e_1}{2m_1}({\bm S}_1\cdot{\bm B})\biggl\{g_1\biggl[(1-\frac{\bm p^2}{2m_1^2})-\frac{e_2[e_1-(e_1+e_2)\mu_1^2]}{24 \pi m_1r}-\frac{e_2[e_2-(e_1+e_2)\mu_2^2]}{12\pi m_2r}\biggr]
\nonumber\\
+(g_1-2)\biggl[\frac{{\bm p}^2}{3m_1^2}
-\frac{e_2[e_1-(e_1+e_2)\mu_1^2]}{24 \pi m_1r}\biggr]\biggr\}.
\end{eqnarray}
\end{widetext}

\noindent
The Hamiltonian for the second constituent has the same form. The leading binding  correction to the $g$-factor is completely described by this Hamiltonian. We calculate its matrix element with the help of the first order perturbation theory between the Schr\"odinger-Coulomb wave functions that are eigenfunctions  of the unperturbed internal Hamiltonian. After simple calculations we obtain the bound state $g$-factors with account of the leading relativistic corrections of order $(Z\alpha)^2$ for $s$-states with the principal quantum number $n$

\begin{widetext}
\begin{eqnarray} \label{g1}
g_1^{bound}=
g_1\biggl[(1-\frac{\mu_2^2e_1^2e_2^2}{2(4\pi)^2 n^2})+\frac{\mu_2e_1e^2_2[e_1-(e_1+e_2)\mu_1^2]}{6 (4\pi)^2 n^2}+\frac{\mu_1e_1e^2_2[e_2-(e_1+e_2)\mu_2^2]}{3(4\pi)^2n^2}\biggr]
\nonumber\\
+(g_1-2)\biggl[\frac{\mu_2^2e_1^2e_2^2}{3(4\pi)^2 n^2}
+\frac{\mu_2e_1e_2^2[e_1-(e_1+e_2)\mu_1^2]}{6 (4\pi)^2 n^2}\biggr],
\end{eqnarray}
\begin{eqnarray} \label{g2}
g_2^{bound}=
g_2\biggl[(1-\frac{\mu_1^2e_1^2e_2^2}{2(4\pi)^2n^2})
+\frac{\mu_1e_1^2e_2[e_2-(e_1+e_2)\mu_2^2]}{6 (4\pi)^2n^2}+\frac{\mu_2e_1^2e_2[e_1-(e_1+e_2)\mu_1^2]}{3(4\pi)^2n^2 }\biggr]
\nonumber\\
+(g_2-2)\biggl[\frac{\mu_1^2e_1^2e_2^2}{3(4\pi)^2n^2}
+\frac{\mu_1e_1^2e_2[e_2-(e_1+e_2)\mu_2^2]}{6 (4\pi)^2n^2 }\biggr].
\end{eqnarray}
\end{widetext}

These results  resolve the discrepancy mentioned in the Introduction in favor of the results in \cite{eg} (see also \cite{pach2008}). The remarkable property of the expressions in \eq{g1} and \eq{g1} is that they are universal for particles of any spin; they depend only on the $g$-factors of free charged particle, not on the magnitude of their spin.  Technically this happened because all terms in the effective two-particle NRQED Hamiltonian in \eq{totham} relevant for calculation of the leading relativistic corrections do not contain spin-dependent coefficients $\lambda_i$, $\bm \Sigma^2_i$. On the other hand analysis of dimensions and spin structure of all terms in the NRQED Lagrangian in \eq{lagr} leads to the conclusion that terms with derivatives of electric fields do not generate contributions to the leading relativistic corrections to the  bound state $g$-factors. Omission of the field derivatives is the basic assumption for validity of the Bargmann-Michel-Telegdi (BMT) equation \cite{bmt,blp}. Hence, the approximation based on the BMT equation \cite{eg} is sufficient for calculation of the the leading nonrecoil relativistic corrections to bound state $g$-factors. Then the leading relativistic corrections are universal because the BMT equation is universal for all spins. For purely electromagnetically interacting particles the free $g$-factors in the BMT equation and in Eqs.~(\ref{g1}-\ref{g2}) are just sums of the QED perturbation series. The BMT equation alone is insufficient for calculation of the leading relativistic recoil corrections and should be amended by the one-photon exchange potential in \eq{trexch}. This is a spin-orbit interaction, it does not depend on magnitude of spins, and it produces universal corrections to bound state $g$-factors.

\begin{acknowledgments}
We are deeply grateful to Peter Mohr and Barry Taylor who attracted our attention to the problem discussed in this note. We greatly appreciate the help, advice, and input of Howard Grotch who participated at the initial stage of this project. This work was supported by the NSF grant PHY--0757928.
\end{acknowledgments}

\end{document}